\documentclass[12pt]{article}
\def\tbar {\overline{t}}
\def\beq {\begin{equation}}
\def\eeq {\end{equation}}
\def\be {\begin{equation}}
\def\ee {\end{equation}}
\def\barr{\begin{array}}
\def\earr{\end{array}}
\def\bea{\begin{eqnarray}}
\def\eea{\end{eqnarray}}
\def\bmath{\begin{displaymath}}
\def\emath{\end{displaymath}}
\def\bq{\begin{quote}}
\def\eq{\end{quote}}
\def\oas{$O(\alpha_s)$}

\def\g5{\gamma_5}
\def\as{\alpha_s}
\def\real{\mathop{\mbox{\rm Re}}\nolimits}
\def\imag{\mathop{\mbox{\rm Im}}\nolimits}

\def\CF{C_{\scriptscriptstyle F}}

\def\c2t{\cos^2\kern-2pt\theta}

\def\s2t{\sin^2\kern-2pt\theta}

\def\Li{\mbox{$\mbox{\rm Li}_2$}}
\def\Frac#1#2{\mbox{$\textstyle{#1\over#2}$}}
\def\half{\Frac{1}{2}}

\def\nn{\nonumber\\}

\def\MZ{M_{\scriptscriptstyle Z}}

\def\I#1{{I}_{#1}}
\def\S#1{{S}_{#1}}

\def\II#1{{\tilde{I}}_{#1}}
\def\SS#1{{\tilde{S}}_{#1}}

\def\sint{\int{dy\,dz\over\sqrt{\vphantom{\big|}(1-y)^2-\xi\,}\:}\kern.1cm}

\def\sxi{\mbox{$\sqrt\xi$}}
\def\Deltab#1{{\bf\bar{\rm\Delta}}^{#1}}
\def\Dsigmap{\Delta\sigma^{(+)}}
\def\Dsigmam{\Delta\sigma^{(-)}}
\def\lapp{\stackrel{<}{~}}

\begin{document}
\vspace*{-2.5cm}
\begin{flushright}
hep-ph/9904319 \\ [-.2cm]
FTUV-97/51 \\ [-.2cm]
IFIC-97/82 \\ [-.2cm]
PRL-TH-97/28 \\ [-.2cm]
ITP-SB-97-67 \\
\end{flushright} 
\begin{center}
\vskip-.25cm
{\Large \bf Longitudinal quark polarization in
            {\boldmath $e^+e^- \to t\overline{t}$} \\[-.43cm]
             and chromoelectric and chromomagnetic \\
             dipole couplings of the top quark}
\vskip.75cm
Saurabh D.~Rindani$^a$ and Michael M.~Tung$^b$
\vskip.4cm
{\it Instituto de F\'{\i}sica Corpuscular, Departament de F\'{\i}sica Te\`orica\\
Universitat de Val\`encia, 46100 Burjassot (Val\`encia), Spain}\\
\vskip.25cm
\it{$^a$Theory Group, Physical Research Laboratory\\
Navrangpura, Ahmedabad 380 009, India}
\vskip.25cm
\it{$^b$Institute for Theoretical Physics\\
State University of New York\\
Stony Brook, NY 11794-3840, U.S.A}
\vskip.5cm
{\bf Abstract}
\end{center}
\vskip-.5cm
The effect of anomalous chromomagnetic ($\mu$) and chromoelectric couplings
($d$) of the gluon to the top quark are considered in $e^+e^- \to 
t\overline{t}$, with unpolarized and longitudinally polarized electron beams.
The total cross section, as well as $t$ and $\tbar$ polarizations are
calculated to order $\alpha_s$ in the presence of the anomalous couplings.
One of the two linear combinations of $t$ and $\tbar$ polarizations is CP
even, while the other is CP odd.
The limits that could be obtained at a typical future linear collider with
an integrated luminosity of 50 fb$^{-1}$ and a total c.m. energy of 500~GeV 
on the
most sensitive CP-even combination of anomalous couplings are estimated as
$-3 \lapp {\rm Re}(\mu) \lapp 2$, for ${\rm Im}(\mu)=0=d$ and 
$\sqrt{{\rm Im}(\mu)^2 + \vert d \vert ^2} \lapp 2.25$ 
for ${\rm Re}(\mu)=0$. There is an improvement by roughly a factor of 2 
at 1000~GeV. 
On the other hand, from 
the CP-odd combination, we derive the possible complementary bounds as
$-3.6 < {\rm Im} (\mu^* d) < 3.6$ for ${\rm Im} (d) = 0$ 
and $-10 < {\rm Im} (d) < 10$ for ${\rm Im} (\mu^* d) = 0$, for a c.m. energy
of 500~GeV. The corresponding limit for 1000~GeV is almost an order of magnitude better for ${\rm Im} (\mu^* d)$, though somewhat worse for ${\rm Im} (d)$.
Results for the c.m. energies 500~GeV and 1000~GeV, if combined, would yield
{\em independent} limits on the two CP-violating parameters of
$-0.8 < {\rm Im} (\mu^* d) < 0.8$ and $-11 < {\rm Im} (d) < 11$. \\[.2cm]
\newpage
\section{Introduction}
The discovery of a heavy top quark, with a mass of $m_t = 175 \pm 6$ GeV
\cite{CDF}, which is far larger than that of all other quarks, opens up the
possibility that the top quark may have properties very different from those of
the other quarks. Observation of these properties might even signal new physics
beyond the standard model. Several efforts in the past few years have gone
into the investigation of the potential of different experiments to study
possible new interactions of the top quark. In particular, possible anomalous
couplings of the top quark to electroweak gauge bosons\footnote{References
to the voluminous literature on this subject can be found, for example, in
\cite{ttbargamma}.} and
to gluons \cite{ttbarg, rizzog} have also been discussed.  Top polarization is
especially useful in such studies \cite{ps,hp,gh,yuan}, because with a mass
around 175 GeV, the top quark decays before it can hadronize \cite{bigi}, and
all spin information is preserved in the decay distributions.

In this paper, we investigate the potential of $e^+e^-$ experiments at a future
linear collider with centre-of-mass (c.m.) energies of 500~GeV or higher, to
study anomalous chromomagnetic and chromoelectric dipole couplings of the top
quark to gluons. So far, a considerable amount of earlier work on the topic of
anomalous gluon couplings has concentrated on hadron colliders. But also
high energy $e^+e^-$ experiments with sufficiently high luminosities would
provide a relatively clean environment to probe the standard model for
anomalous gluon couplings. While earlier efforts in the context of $e^+e^-$
colliders are mainly based on an analysis of the gluon distribution
\cite{rizzog} in $e^+e^-\to t\tbar g$, we look at the possible
information that could be obtained from studying the total cross section,
and the polarization of $t$ and $\tbar$  separately. This has the advantage
over $t$ and $\tbar$ spin correlations that, because the polarization of only
one of $t$ and $\tbar$ is analyzed by means of a definite decay channel,
the other is free to decay into any channel. This leads to much better
statistics compared to the case when $t-\tbar$ spin correlations are
considered, where definite $t$ and $\tbar$ channels have to be used as
analyzers.

We find that there are three independent quantities, viz., the cross section,
and the CP-even and CP-odd linear combinations of the $t$ and $\tbar$
polarizations, which can be used to probe separately the CP-even chromomagnetic
dipole moment coupling and the CP-odd chromoelectric dipole coupling. Of these,
the CP-odd combination provides the most sensitive probe of the imaginary parts
of the anomalous couplings.

We obtain here the 90\% confidence level (C.L.) limits that would be possible
at a typical $e^+e^-$ linear collider, with integrated luminosity 50 fb$^{-1}$.
We have considered two possible centre-of-mass (c.m.) energies, viz., 500~GeV
and 1000~GeV.
We have also considered the effect of beam polarization on the sensitivity
of the measurements. 

Our results on the constraints on the CP-violating
chromoelectric dipole moments were presented in Ref.~\cite{RT} . We include in
this paper also the limits that would be obtainable on CP-conserving
chromomagnetic dipole moments, and combinations of couplings.

The paper is organized as follows: In Section 2, we introduce the effective
action for the general massive anomalous $q\bar{q}g$ vertex. The subsequent
discussion presents the full framework necessary for obtaining the
analytical extensions that modify the standard QCD one-loop predictions
for quark-antiquark production. We then give explicit combinations of the
total polarized cross sections which are sensitive to the chromomagnetic and
chromoelectric dipole moments, respectively. Next, Section 3 focuses
on the numerical estimates of these observables for various kinematical
regions readily accessible at a future $e^+ e^-$ collider. The final
Section 4 is the Conclusion.

\section{Calculation of cross section and top polarization with anomalous
         couplings}
An effective $t\tbar g$ vertex can be written in the form
\beq
\Gamma^a_{t\tbar g} = - g_s T^a \left[  \gamma^{\mu}\epsilon_{\mu} + 
 \frac{i\mu}
{m_t} \sigma^{\mu\nu}q_{\mu}\epsilon_{\nu}
- \frac{d}{m_t} \sigma^{\mu\nu}\gamma_5 q_{\mu}\epsilon_{\nu} \right],
\label{effL}
\eeq
where
\beq
T^a = \Frac{1}{2} \lambda^a; 
\eeq
$\lambda^a$ being the $SU(3)$ Gell-Mann
matrices, and $q_{\mu}$ and $\epsilon_{\mu}$ are respectively the momentum and
polarization four-vectors of the gluon. 
This is the most general Lorentz- and colour-invariant trilinear
coupling (additional quadrilinear terms are needed for local
colour invariance, but we do not need them here). The $\mu$ and $d$ terms
are the chromomagnetic and chromoelectric dipole terms, respectively.
These dipole couplings $\mu$ and $d$ are in fact momentum-dependent form
factors, and complex in general. They parameterize the effects arising at
the loop level, which in principle could arise from both, interactions
within the Standard Model (SM) and possible new interactions. If the
measured values of the corresponding form factors deviate from the
theoretical predictions of the conventional SM, this would indicate
the presence of additional, new physics. Note that a possible
experimental determination of non-vanishing $\imag(\mu)$ [$\imag(d\,)$]
does not necessarily imply the existence of CP-conserving [CP-violating]
new physics.

We use Eq.~({\ref{effL}) to calculate the $t\tbar$ total cross section and the
$t$ and $\tbar$ polarizations  to order $\alpha_s\equiv g_s^2/(4\pi)$.
The extra diagrams contributing to this order are shown in Fig.~1, where the
large dots represent anomalous couplings. The anomalous couplings enter in the
amplitudes for soft and collinear gluon emissions. The infrared divergences in
the amplitudes for soft gluon emission and in the virtual gluon corrections
cancel as in standard QCD, since the anomalous terms vanish in the infrared
limit. Moreover, we do not include anomalous couplings in the loop diagrams.
The latter are included merely to regulate the infrared divergences. We thus
study how anomalous couplings at tree-level would modify standard QCD
predictions at order $\alpha_s$.

For heavy-quark production, the standard QCD one-loop corrections to the total
cross section and the longitudinal spin polarization~\cite{kpt} were calculated
in closed analytic form before. Those results have been extended here to
the case when anomalous couplings are present.

The total unpolarized production rate is given only in terms of the $VV$ and
$AA$ parity-parity combinations for the Born contributions:
\be
\sigma_{Born}\left(e^+ e^-\to\gamma,Z\to q\bar{q}\right) =
\Frac{1}{2}v(3-v^2)\sigma^{VV}+v^3\sigma^{AA},
\ee
where the mass parameters are $v=\sqrt{1-\xi\,}$ and $\xi=4m_q^2/s$.
The $O(\as)$ unpolarized case has the cross section
\be
\sigma\left(e^+ e^-\to\gamma,Z\to q\bar{q}\right) =
\Frac{1}{2}v(3-v^2)\sigma^{VV}c^{VV}+v^3\sigma^{AA}c^{AA},
\ee
where the $VV$ and $AA$ factors that
multiply with the appropriate Born terms are given below.
\be
c^{VV} = 1+{\as\over2\pi}\CF\left[\,\tilde{\Gamma}-v{\xi\over2+\xi}\ln\left(
         {1+v\over1-v}\right)-{4\over v}\I{2}-{\xi\over v}\II{3}+
         {4\over v(2+\xi)}+{2-\xi\over v}\II{5}+\Delta_0^{VV}\,\right].
\label{cvv}
\ee
In this equation, the contribution from the virtual gluon loop is denoted as
\bea
\tilde{\Gamma} &=&
  \left[\,2-{1+v^2\over v}\ln\left({1+v\over1-v}\right)\right]\,
  \ln\left(\Frac{1}{4}\xi\right)+{1+v^2\over v}\left[\,
  \Li\left(-{2v\over1-v}\right)-\Li\left({2v\over1+v}\right)+
  \pi^2\,\right] \nn
  &&+3v\ln\left({1+v\over1-v}\right)-4.
\eea
Here, the $q\bar{q}g$ phase-space integrals are abbreviated by $\I{i}$,
and $\II{i}$ specify the results after the (soft) IR divergences have canceled.
The explicit analytical expressions for these phase-space integrals may be
found in \cite{tbp}. The additional component stemming from the anomalous
gluon bremsstrahlung is given by
\be
\Delta_0^{VV} = {8\over(2+\xi)v}\left[\,\real(\mu)(\I{1}+\I{4})+{2\over\xi}
  \left(|\mu|^2+|d|^2\right)(\I{1}-2\I{8})\,\right].
\ee
For the $AA$ contribution we find:
\be
c^{AA} = 1+{\as\over2\pi}\CF\left[\,\tilde{\Gamma}+2{\xi\over v}\ln\left(
         {1+v\over1-v}\right)+{\xi\over v^3}\I{1}-{4\over v}\I{2}-
         {\xi\over v}\II{3}+{2+\xi\over v^3}\I{4}-{2-\xi\over v}\II{5}
         +\Delta_0^{AA}\,\right],
\label{caa}
\ee
with the following anomalous part
\bea
\Delta_0^{AA} &=&
   {2\over v^3}\left[\,\real(\mu)\Big\{-(4-\xi)\I{1}+(2+\xi)\I{4}\Big\}
   +\left(|\mu|^2+|d|^2\right)\left\{\left({4\over\xi}+\xi-6\right)\I{1}+
   \xi\I{4}
   \right.\right.\nn&&\left.\left.
   -{4\over\xi}(2-\xi)\I{8}+{4\over\xi}\I{9}\right\}\,\right].
\eea
The remaining $V\!A$ and $AV$ parts are identical and only contribute to the
spin-dependent cross section. In the absence of anomalous couplings $(\mu=d=0)$, the cross section for longitudinally polarized quarks of helicity $\pm \half$
is given by
\be
\sigma\left(e^+ e^-\to\gamma,Z\to q(\lambda_\pm)\,\bar{q}\right) =
\Frac{1}{2}v(3-v^2)\sigma^{VV}c^{VV}+v^3\sigma^{AA}c^{AA}
\pm v^2\sigma^{V\!A}_S c_\pm^{VA}.
\ee
The multiplication factors $c^{ij}$ are expressed in terms of phase-space
integrals of type $\S{i}$:
\be
c^{V\!A,AV}_\pm =
1+{\as\over2\pi}\CF\left[\,\tilde{\Gamma}+{\xi\over v}\ln\left({1+v\over1-v}
\right)+\Delta^{V\!A,AV}_{\mu=d=0}\,\right],
\label{cva}
\ee
where
\bea
\Delta^{V\!A,AV}_{\mu=d=0} &=&
\Frac{1}{2}\Big[\,(4-\xi)\S{1}-(4-5\xi)\S{2}-2(4-3\xi)\S{4}-
\xi(1-\xi)(\SS{3}+\SS{5})+\xi(\S{6}-\S{7}) \nn
&& -2\S{8}+(2-\xi)\S{9}+(6-\xi)\S{10}-2\S{11}+
2(1-\xi)(2-\xi)\S{12}\,\Big].
\eea
The full analytic forms of all relevant $S$ integrals are too lengthy to
be exhibited here. Most of them are compiled in Ref.~\cite{tbp}, except
for the four additional integrals $S_{14}$, $S_{15}$, $S_{16}$, and
$S_{17}$, which are discussed in more detail in the appendix.

Including spins for the quark or the antiquark introduces additional
spin-flip terms in the $O(\as)$ $c$ factors given in Eqs.~(\ref{cvv}),
(\ref{caa}) and (\ref{cva}).
For longitudinal quark polarization we find
\be
c^{ij}_\pm =
\Frac{1}{2}\left[\,c^{ij}\pm{\as\over2\pi}\CF\Delta^{ij}_S\,\right].
\ee
The individual parity-parity combinations are
\bea
\label{sf1}
(2+\xi)v\,\Delta^{VV}_S &=&
 8\imag\left(\mu^*d\right)\Bigg[\,
-2\left(1-{2\over\xi}\right)\S{1}+\left(1-{4\over\xi}\right)\S{8}-\S{9}-
\S{10}+\S{11} \nn
&&\hskip2.4cm+2\left(1-{4\over\xi}\right)\S{13}+{4\over\xi}(\S{15}+\S{17})\,
\Bigg] \nn
&&+\imag(d)\Bigg[\,8(1-\xi)\S{1}-\Big\{8-3\xi(2-\xi)\Big\}\S{2}+
\Big\{8+\xi(2-3\xi)\Big\}\S{4} \nn
&&\hskip1.75cm-\xi(2+\xi)\S{6}-4\S{8}+4(1-\xi)\S{9}-4(3+\xi)\S{10}+4\S{11} \nn
&&\hskip1.75cm+\xi(2-\xi)\S{14}
\,\Bigg], \\[.5cm]
\label{sf2}
v^3 \Delta^{AA}_S &=&
4\imag\left(\mu^*d\right)\Bigg[\,
{2\over\xi}(1-\xi)(2-\xi)\S{1}+\left(5-{4\over\xi}\right)\S{8}-
(1-\xi)(\S{9}+\S{10})+\S{11} \nn
&&+2\left(3-{4\over\xi}\right)\S{13}-
2\left(1-{2\over\xi}\right)\S{15}-{4\over\xi}\S{16}-
2\left(1-{4\over\xi}\right)\S{17}\,\Bigg] \nn
&&+\imag(d)\Bigg[\,
2\xi\S{1}-(1-\xi)(4-\xi)(\S{2}-\S{4})-\xi(1-\xi)\S{6}-(2-\xi)\S{8} \nn
&&+(2-3\xi)\S{9}-(6-5\xi)\S{10}+(2+\xi)(\S{11}+2\S{13})+\xi(1-\xi)\S{14}
\Bigg],\\[.5cm]
v^2 \Delta^{V\!A,AV}_S &=&
\real(\mu)\Bigg[\,
-\xi(\S{1}+\S{7})-2\S{8}+(2-\xi)(\S{9}+\S{10})-2\S{11}\,\Bigg] \nn
&&\pm2\,i\imag(\mu)\Bigg[\,
(2-\xi)\S{1}-\xi\S{10}-2\S{13}\,\Bigg] \\
&&+\left(|\mu|^2+|d|^2\right)\Bigg[\,-(4-\xi)\S{1}-\xi\S{7}+4\S{8}+\xi\S{9}+
(4-\xi)\S{10}-4\S{11}\,\Bigg]. \nonumber
\eea
Using charge conjugation in the final state, one can readily obtain the
corresponding expressions for (longitudinal) antiquark polarization. In the
following, we denote the antiquark results by an additional bar, {\it i.e.\/}
$\Deltab{ij}_S$:
\bea
\Deltab{VV}_S &=& \Delta^{VV}_S, \\
\Deltab{AA}_S &=& \Delta^{AA}_S,
\eea
where the following identities hold
\bea
\Deltab{V\!A} &=& -\Delta^{V\!A}_S\ =\ \left(\Deltab{AV}_S\right)^*, \\
\hbox{with}\qquad \Deltab{AV}_S &=& -\Delta^{AV}_S.
\eea

Considering the above expressions, we can construct the following
combinations of polarization asymmetries of $t$ and $\tbar$,
\beq
\Delta\sigma^{(+)} = \Frac{1}{2} \Big[\,
\sigma(\uparrow ) - \sigma(\downarrow)
-
\overline{\sigma}(\uparrow) + \overline{\sigma}(\downarrow)
\,\Big],
\eeq
\beq
\Delta\sigma^{(-)} = \Frac{1}{2} \Big[\,
\sigma(\uparrow) - \sigma(\downarrow)
+
\overline{\sigma}(\uparrow) - \overline{\sigma}(\downarrow)
\,\Big],
\eeq
where $\sigma (\uparrow)$, $\overline{\sigma} (\uparrow)$
refer respectively
to the cross sections for top and antitop with positive helicity, and
$\sigma (\downarrow)$, $\overline{\sigma} (\downarrow)$
are the same quantities with negative helicity. Of these, $\Delta\sigma^{(+)}$
is CP even and $\Delta\sigma^{(-)}$ is CP odd. This is obvious from the fact
that under C, $\sigma$ and $\overline{\sigma}$ get interchanged, while under
P, the helicities of both $t$ and $\tbar$ get flipped. Consequently, $\sigma$
and $\Delta\sigma^{(+)}$, both nonzero in standard QCD, receive contributions
from combinations of anomalous couplings which are CP even, viz., Im$(\mu)^2 +
\vert d\vert ^2$ and Re$(\mu)$. On the other hand, $\Delta\sigma^{(-)}$
vanishes in standard QCD, and in the presence of anomalous couplings it
depends only on the CP-odd variables Im$(\mu^*d)$ and Im$(d)$. That
$\Delta\sigma^{(-)}$ depends on the imaginary parts rather than the real parts
of a combination of couplings follows from the fact that it is even under
naive time reversal T$_{\rm N}$, i.e., reversal of all momenta,
without change in helicities, and without interchange of initial and
final states (as would have been required by genuine time reversal). As a
consequence, it is odd under CPT$_{\rm N}$, and imaginary parts of couplings
have to appear in order to avoid conflict with the CPT theorem.

\section{Numerical results}
Figs.~2a and 2b show the dependence of $\Dsigmap$ on
$\sqrt{{\rm Im}(\mu)^2+\vert d\vert^2}$ and Re$(\mu)$ at $\sqrt{s}=500$ GeV
and 1000 GeV, respectively. Figs.~3a and 3b depict $\Dsigmam$ plotted against
Im$(\mu^*d)$ and Im$(d)$.

We would now like to see how sensitive experiments at a future linear collider
would be to the anomalous quantities $\mu$ and $d$. Since these are determined
through $\Dsigmap$ and $\Dsigmam$, we should know how well these latter
quantities can be measured experimentally. The quantity $\sigma(\uparrow)-
\sigma(\downarrow)$ is simply the polarization $P_t$ of the top quark in
the production process, and it can be determined by looking at the angular
distribution of the decay products of the top. For example, in the decay mode
$t\rightarrow W^+b$, the angular distribution of the $b$ quark in the $t$ quark
rest frame is
\beq
\frac{1}{\Gamma}\frac{d\Gamma}{d\Omega_{b}} = \frac{1}{4\pi}
	\Big(1+P_t\,\alpha\,\hat{p}_b\cdot\vec{s}\,\Big),
\label{tpol}
\eeq
where $\vec{p}_b$ is the $b$-quark momentum and $\vec{s}$ is the top
spin. This angular distribution can be used to determine $P_t$.
In Eq.~(\ref{tpol}), the parameter $\alpha$ ($\vert\alpha\vert\leq 1$) is
a constant known as the {\it analyzing power} for the decay channel.
In this particular case, $\alpha$ is given by
\beq
\alpha=\frac{m_t^2-2m_W^2}{m_t^2+2m_W^2}\ .
\eeq
There is an analogous expression for the decay
$\overline{t}\rightarrow W^-\overline{b}$.
If the angular distribution of a lepton or a jet arising from $W$ decay is
used to determine $P_t$, the corresponding analyzing power would be different. The
efficiency with which the top or antitop polarization can be measured will
depend not only on the analyzing power of the channel, but also on the detection
efficiency of the observed particles, like the $b$ quark. While the tagging
efficiency of the $b$ would be much better at $e^+e^-$ colliders than at hadron
colliders, because of the lower hadronic activity, it would nevertheless depend
on the degree to which backgrounds are understood and eliminated. A more
detailed analysis in the context of specific experimental conditions would be
needed to obtain the overall top polarization detection efficiency, which we
parameterize as $\epsilon$ in what follows.

We use our expressions to obtain simultaneous 90\% confidence level (CL)
limits that could be obtained at a future linear collider with an integrated
luminosity of 50 fb$^{-1}$. We do this by equating the magnitude of the
difference between the values for a quantity with and without anomalous
couplings to 2.15 times the statistical error expected. Thus, the limiting
values of Im$(\mu)^2 + \vert d\vert ^2$ and Re$(\mu)$ for an integrated
luminosity $L$ and a top detection efficiency of $\epsilon$ are obtained from
\beq
\epsilon\, L\,\vert\,\sigma(\mu,d)-\sigma_{\scriptscriptstyle SM}\,\vert
= 2.15\;\sqrt{L\,\sigma_{\scriptscriptstyle SM}\,}
\label{lims}
\eeq
and
\beq
\epsilon\, L\,\left\vert\,\Dsigmap(\mu,d)-\Dsigmap_{\scriptscriptstyle SM}
\,\right\vert = 2.15 \,\sqrt{L\,\left\vert\,\sigma_{\scriptscriptstyle SM}
(\uparrow)-\overline{\sigma}_{\scriptscriptstyle SM}(\uparrow)\,\right\vert\:},
\label{lim+}
\eeq
whereas the limiting values of Im$(\mu^* d)$ and Im$(d)$ are obtained from
\beq
\epsilon\, L\,\left\vert\,\Dsigmam(\mu,d)-\Dsigmam_{\scriptscriptstyle SM}
\,\right\vert = 2.15 \,\sqrt{L\,\left\vert\,\sigma_{\scriptscriptstyle SM}
(\uparrow)+\overline{\sigma}_{\scriptscriptstyle SM}(\uparrow)\,\right\vert\:}.
\label{lim-}
\eeq
In the above expressions, the subscript ``SM'' denotes the value expected in the
standard model, with $\mu=d=0$. We use $L=50$ fb$^{-1}$ and $\epsilon = 0.1$
in our numerical estimates. 
The value of $\epsilon$ depends on the details of the detector, as well as the
kinematical cuts employed. The value we use is only representative, and it
would be easy to obtain limits for any other value of $\epsilon$ by appropriate
scaling, because of the simple dependence on $\epsilon$ in Eqs.~(\ref{lims}),
(\ref{lim-}), and (\ref{lim+}).

For the running of the strong coupling, we choose
$\as^{(5)}(\MZ)=0.118$ (with $\MZ=91.178$~GeV) in the modified minimal
subtraction scheme and use the appropriate conditions to match for six active
flavours\footnote{It is common to indicate the number of active flavours as
superscript of $\as$. For practical purposes one usually selects bottom
production as reference. Here, our choice for $\as^{(5)}(\MZ)$ translates to
$\as^{(6)}(M_t=172.1{\rm GeV})=0.10811$}.

Eqs.~(\ref{lims}) and (\ref{lim+}) are used to obtain contours in the plane
of $\sqrt{{\rm Im}(\mu)^2+\vert d\vert^2}$ and Re$(\mu)$. The contours from
(\ref{lims}) are shown in Figs.~4a and 4b for $\sqrt{s}=500$~GeV and 1000~GeV,
respectively. The contours obtained from (\ref{lim+}) are shown in Figs.~5a
and 5b. Eq.~(\ref{lim-}) gives contours in the plane of Im$(\mu^*d)$ and
Im$(d)$, which are displayed in Figs.~6a and 6b. For different $e^-$
longitudinal beam polarizations $P_-$, the corresponding contours are presented
in Figs.~4--6. In Figs.~4 and 5, the allowed regions are the ones shown below
the respective contours. In Figs.~6a,b, the allowed regions are bands lying
between the upper and lower straight lines.

In principle, the measurement of two independent quantities $\sigma$ and
$\Dsigmap$ could have given independent limits on Im$(\mu)^2+\vert d\vert^2$
and Re$(\mu)$. However, a comparison of Figs.~4 and 5 shows that this
possibility is not realized. A superposition of Figs.~4a,b and 5a,b is shown
in Fig.~7a. It can be seen that the allowed region coming from $\sigma$ lies
entirely within the allowed region from $\Dsigmap$, with no intersection
between the two. Thus $\sigma$ is more sensitive to the anomalous CP-even
couplings as compared to $\Dsigmap$. Moreover, each contour for
$\sqrt{s}=1000$~GeV lies within the corresponding contour for
$\sqrt{s}=500$~GeV, showing a uniform increase in sensitivity with c.m.\
energy.

The other conclusion that can be drawn from Figs.~4 and 5 is that a large
left-handed polarization leads to increase in sensitivity.

The best limits obtainable from $\sigma$ at $\sqrt{s}=500$ GeV are $-3\lapp
{\rm Re}(\mu) \lapp 2$ for Im$(\mu)=0=d$, and $\sqrt{{\rm Im}(\mu)^2 + \vert
d \vert ^2} \lapp 2.25$ for Re$(\mu)= 0 $. These limits are improved by about
a factor 2 in going to $\sqrt{s}=1000$~GeV for the same integrated luminosity.

In the case of CP-odd combinations of the couplings Im$(\mu^*d)$ and Im$(d)$,
there is only one measurable CP-odd quantity $\Dsigmam$ at each c.m.\ energy,
and therefore independent limits on the two combinations are not possible.
Fig.~6a shows that for $\sqrt{s}=500$ GeV, the best limits which can
be obtained are for $P=-1$, viz.,
$-3.6 < {\rm Im} (\mu^* d) < 3.6$ for ${\rm Im} (d) = 0$ 
and $-10 < {\rm Im} (d) < 10$ for ${\rm Im} (\mu^* d) = 0$. The corresponding
limits for $\sqrt{s}=1000$ GeV, as seen from Fig.~6b, are
$-0.4 < {\rm Im} (\mu^* d) < 0.4$ for ${\rm Im} (d) = 0$ 
and $-20 < {\rm Im} (d) < 20$ for ${\rm Im} (\mu^* d) = 0$. 
However, if a measurement of $\Dsigmam$ is made at two c.m.\ energies, a
relatively narrow allowed range can be obtained, allowing {\em independent}
limits
to be placed on both Im$(\mu^*d)$ and Im$(d)$. This is demonstrated in Fig.~7b. In fact, the improvement in the limit on Im$(\mu^*d)$ in going from
$\sqrt{s}=500$~GeV to $\sqrt{s}=1000$~GeV is considerable. The possible
limits are
\beq
-0.8 < {\rm Im} (\mu^* d) < 0.8,
-11 < {\rm Im} (d) < 11.
\eeq
These limits may be compared with the limits obtainable from gluon jet energy
distribution in $e^+e^- \to t\tbar g$ \cite{rizzog}. While our proposal
for the CP-even case seems to fare worse,  for the CP-odd case, our proposal
can be competitive. It should however be emphasized that in the case of the
CP-odd couplings, we are proposing the measurement of a genuinely CP-violating
quantity, whereas the analysis in \cite{rizzog} is merely based on the energy
spectrum resulting from both CP-odd and CP-even couplings.
In case of $\Dsigmam$, the dependence on $e^-$ beam
polarization is rather mild.

\section{Conclusions}
It is worthwhile noting that we have used a rather conservative value of 
$\epsilon = 0.1$ for top detection and polarization analysis. A better efficiency
would lead to an improvement in the limits, as would a higher luminosity.

We have not considered the effect of initial-state radiation in this work. We
have also ignored possible effects of collinear gluon emission from one of the
decay products of $t$ or $\tbar$.  A complete analysis should indeed
incorporate these effects, as well as a study of $t$ and $\tbar$ decay
distributions which can be used to measure the polarizations.
However, we do not expect our conclusions
to change drastically when these effects are taken into account.

In summary, we have examined the capability of total cross section and single
quark polarization in $e^+e^- \to t\tbar$ to measure or put limits on
anomalous chromomagnetic and chromoelectric dipole couplings. While the total
cross section measurements can give, for the luminosities assumed, limits of
order 1 on the CP-even couplings, only the CP-violating combination of top and
antitop polarizations is sensitive to anomalous couplings, and can yield a
limit of the order of 1 on a CP-odd combination of anomalous couplings.
\vskip.75cm
{\bf Acknowledgments.} We are both grateful for the kind hospitality and
stimulating scientific atmosphere we enjoyed at the Departament de F\'{\i} sica
Te\`orica during the principal stages of this work. We thank Bar-Shalom
Shaouly for pointing out the update of Figures~6. This work has been
supported by the DGICYT under Grants Ns.\  PB95-1077 and SAB95-0175, as well
as by the TMR network ERBFMRXCT960090 of the European Union (S.D.R.).
M.M.T.\ acknowledges support by CICYT Grant AEN-96/1718 and the Max-Kade
Foundation, New York, NY.
\newpage
\setcounter{equation}{0}
\renewcommand\theequation{A.\arabic{equation}}
\section*{Appendix: Additional Phase-Space Integrals}
The complicated three-body phase-space for the process 
$e^+ e^-\to q(\uparrow)\bar{q}g$ is best solved analytically by
using the kinematic variables $y=1-p_1\cdot q/q^2$ and $z=1-p_2\cdot q/q^2$,
which are natural dimensionless parameters referring to the quark
and antiquark energies (including the radiated gluon) in the centre-of-mass
system. Then, the relevant phase-space integrands all result in simple
rational functions containing polynomials in $y$ and $z$, and the
corresponding integration boundary is described by the symmetric solution
of a $(y,z)$-biquadratic form~\cite{kpt,tbp}.

Apart from the detailed integral list of Ref.~\cite{tbp}, in this particular
calculation four new integrals emerged. For completeness, we give here
the following full analytical results:

\bea
S_{14} &=& \sint {z\over y^2} \nonumber \\[.25cm]
       &=& {2\over\xi}-{2+\sxi\over2(2-\sxi)}-\ln(2-\sxi)+\Frac{1}{2}
           \ln\xi-\Frac{1}{2}, \\
           \nn 
S_{15} &=& \sint\;y^2 \nonumber \\[.25cm]
       &=& \Frac{1}{32}\xi^3\Big[\Frac{1}{2}\ln\xi-\ln(2-\sxi)\Big]-
           \sxi\left(4+\Frac{1}{3}\xi+\Frac{1}{4}\xi^2\right) \nn
       & & +\Frac{1}{8}\left(7-\half\xi\right)\xi+\Frac{1}{3}, \\
           \nn
S_{16} &=& \sint\;y^2z \nonumber \\[.25cm]
       &=& -{\xi^4(4-\xi)\over512(2-\sxi)^2}+
           \Frac{1}{16}\left(\Frac{3}{16}\xi-1\right)\ln(2-\sxi)
           -\Frac{3}{8}\xi\left(1+\Frac{3}{8}\xi\right) \nn
       & & +\Frac{1}{512}\xi^3\Big[4-\xi+(16-3\xi)\ln\xi\Big]
           +\Frac{1}{4}\left(\Frac{7}{3}-
           \Frac{1}{2}\xi+\Frac{1}{16}\xi^2\right)+\Frac{1}{24}, \\
           \nn
S_{17} &=& \sint\;y\,z \nonumber \\[.25cm]
       &=& \Frac{1}{32}\xi^2(6-\xi)\Big[-\Frac{1}{2}\ln\xi+\ln(2-\sxi)\Big]
           +\Frac{1}{128}(4-\xi){\xi^3\over(2-\sxi)^2} \nn
       & & +\Frac{1}{4}\xi^{3\over2}\left(\Frac{5}{3}-\Frac{1}{8}\xi\right)
           -\half\xi\left(1-\Frac{1}{64}\xi^2\right)+\Frac{1}{12},
\eea
where the usual mass parameters are $v=\sqrt{1-\xi}$ and $\xi=4m_q^2/s$.

Note that there are no soft divergences in these particular integrals.
The collinear divergences contained in these integrals arise from the
massless character of the fermion field, and are easily identified
by observing the limit $\xi\to 0$. In our full analytical expressions
for the polarized cross sections with $VV$ and $AA$ parity-parity
combinations, integral $S_{14}$ is multiplied by the mass factor $\xi$,
viz. Eqs.~(\ref{sf1}) and (\ref{sf2}). This produces additional finite
contributions, which originate from a collinear helicity-flip with
anomalous chromoelectric couplings, and are absent in a naive massless
model.
\newpage
\newpage
\thispagestyle{empty}
\centerline{\bf\Large Figure Captions}
\vskip1cm
\newcounter{fig}
\begin{list}{
   \bf Fig.~\arabic{fig}:\ }{
         \usecounter{fig}
         \labelwidth1.6cm
         \leftmargin2cm
         \labelsep0.4cm
         \itemsep0ex plus0.2ex
        }
\item Additional Feynman diagrams contributing to
      $\sigma\left(e^+e^-\to\gamma,Z\to t\bar{t}\right)$
      that account for anomalous gluon couplings at \oas.
      The large dots represent anomalous $t\bar{t}g$ insertions
      according to the effective action Eq.~(1).

\item Surface plots displaying the dependence of the polarization asymmetry
      $\Dsigmap$ on $\sqrt{{\rm Im}(\mu)^2+|d|^2}$ and Re$(\mu)$ with initial
      electron beam polarization $P_-=-1$ and c.m.\ energies {\bf(a)}
      $\sqrt{s}=500$~GeV, {\bf(b)} $\sqrt{s}=1000$~GeV.

\item $\Dsigmam$ surface plots showing the linear dependence on Im$(\mu^*d)$
      and Im$(d)$ for {\bf(a)} $\sqrt{s}=500$~GeV and
      {\bf(b)} $\sqrt{s}=1000$~GeV with longitudinal beam polarization
      $P_-=-1$.

\item Contour plots showing the allowed regions for
      $\sigma_{\scriptscriptstyle SM}$ with 90\% confidence level (integrated
      luminosity $L=50$~fb$^{-1}$ and top detection efficiency $\epsilon=0.1$).
      Representative c.m.\ energies are {\bf(a)} $\sqrt{s}=500$~GeV and
      {\bf(b)} $\sqrt{s}=1000$~GeV for various longitudinal electron
      polarizations.

\item $\Dsigmap$ contour plots with 90\% confidence level at c.m.\ energies:
      {\bf(a)} $\sqrt{s}=500$~GeV and {\bf(b)} $\sqrt{s}=1000$~GeV.

\item $\Dsigmam$ contour plots with 90\% confidence level at c.m.\ energies:
      {\bf(a)} $\sqrt{s}=500$~GeV and {\bf(b)} $\sqrt{s}=1000$~GeV.

\item {\bf(a)} Superposition of Figs.~4 and 5, displaying combined allowed
      regions for both $\sigma_{\scriptscriptstyle SM}$ and $\Dsigmap$
      polarization asymmetries. {\bf(b)} Intersecting area resulting from
      two independent $\Dsigmam$ measurements at $\sqrt{s}=500$~GeV, 1000~GeV.
\end{list}
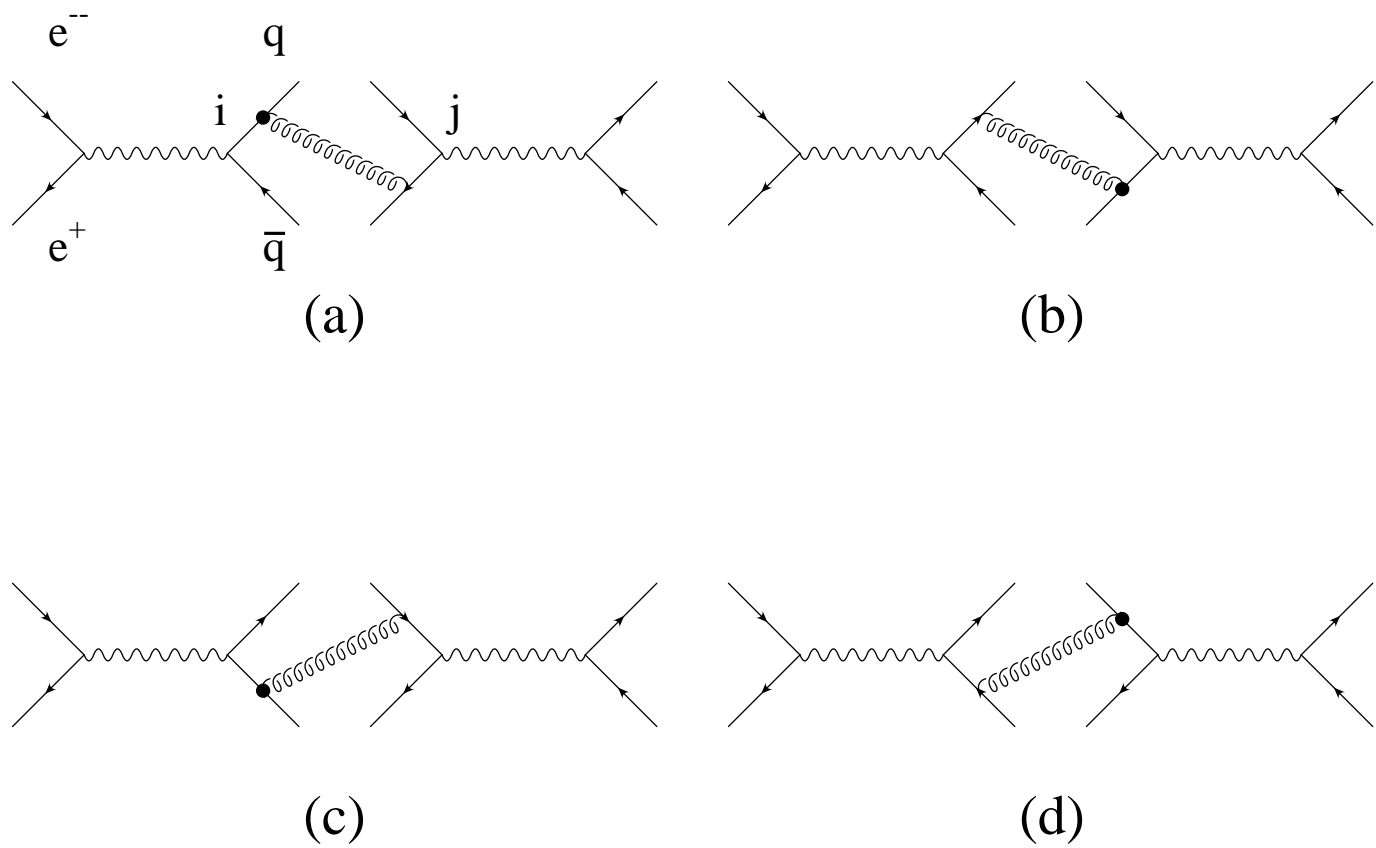
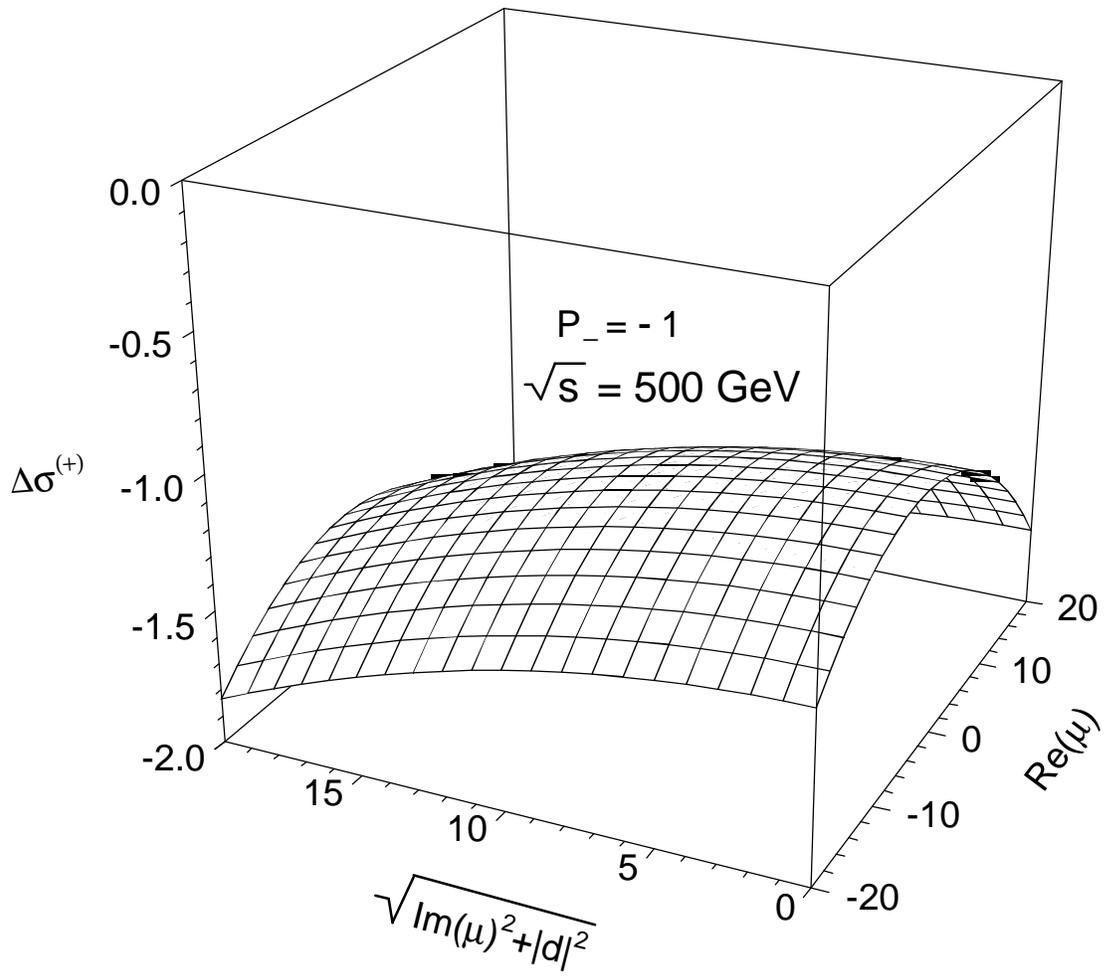
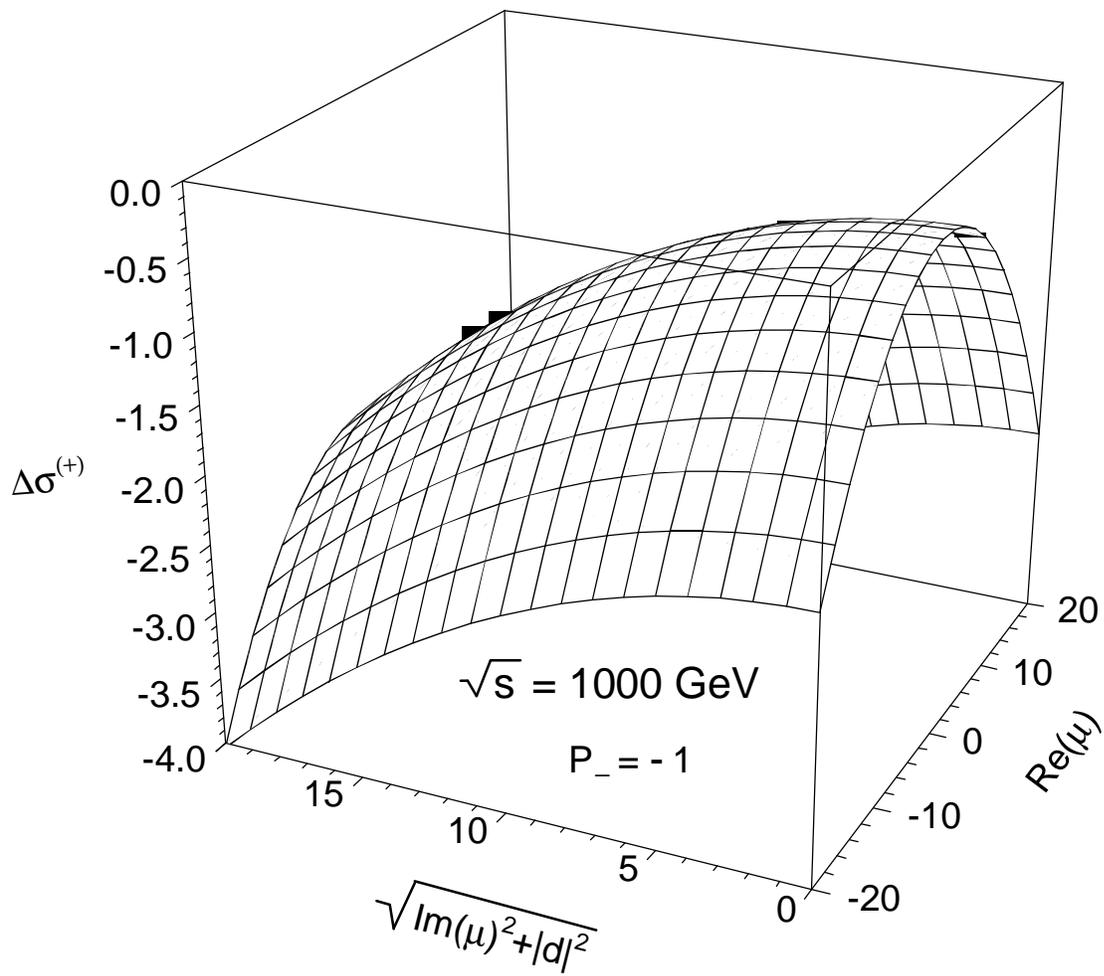
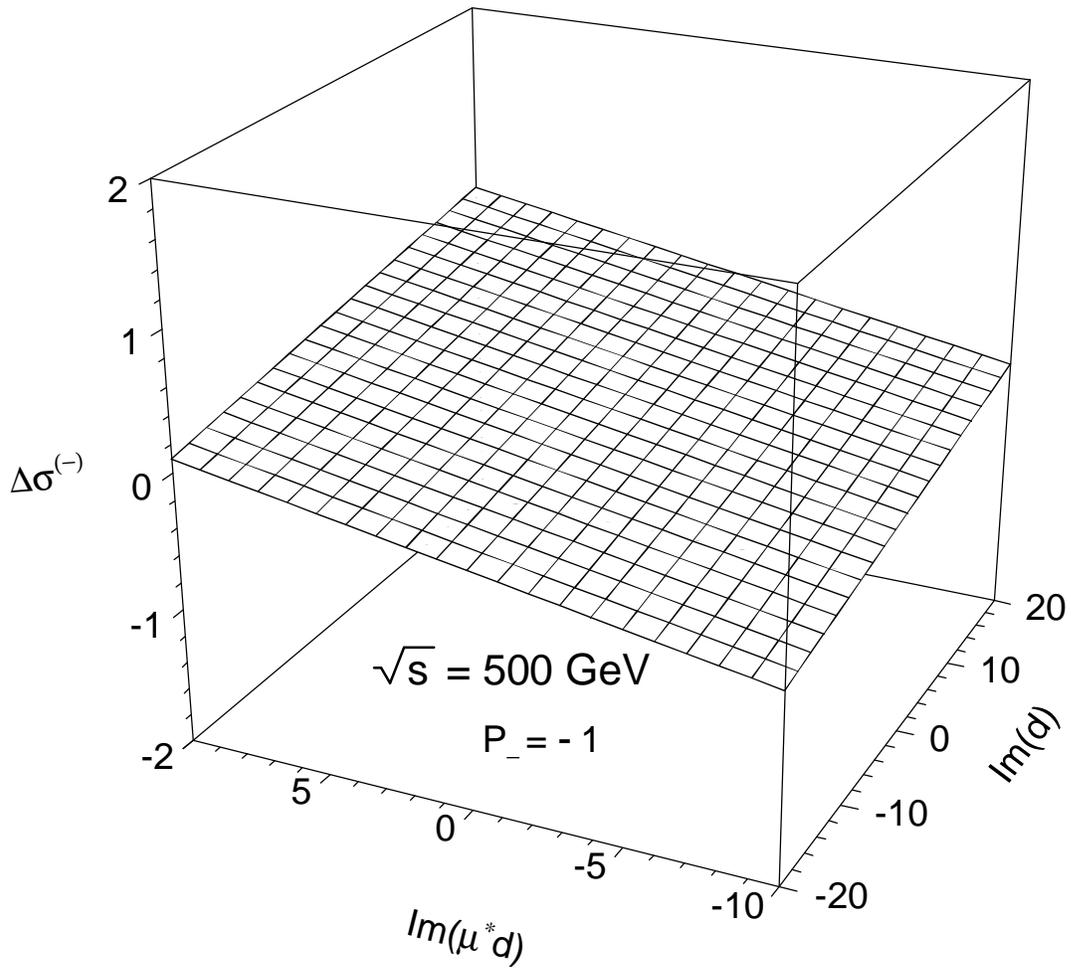
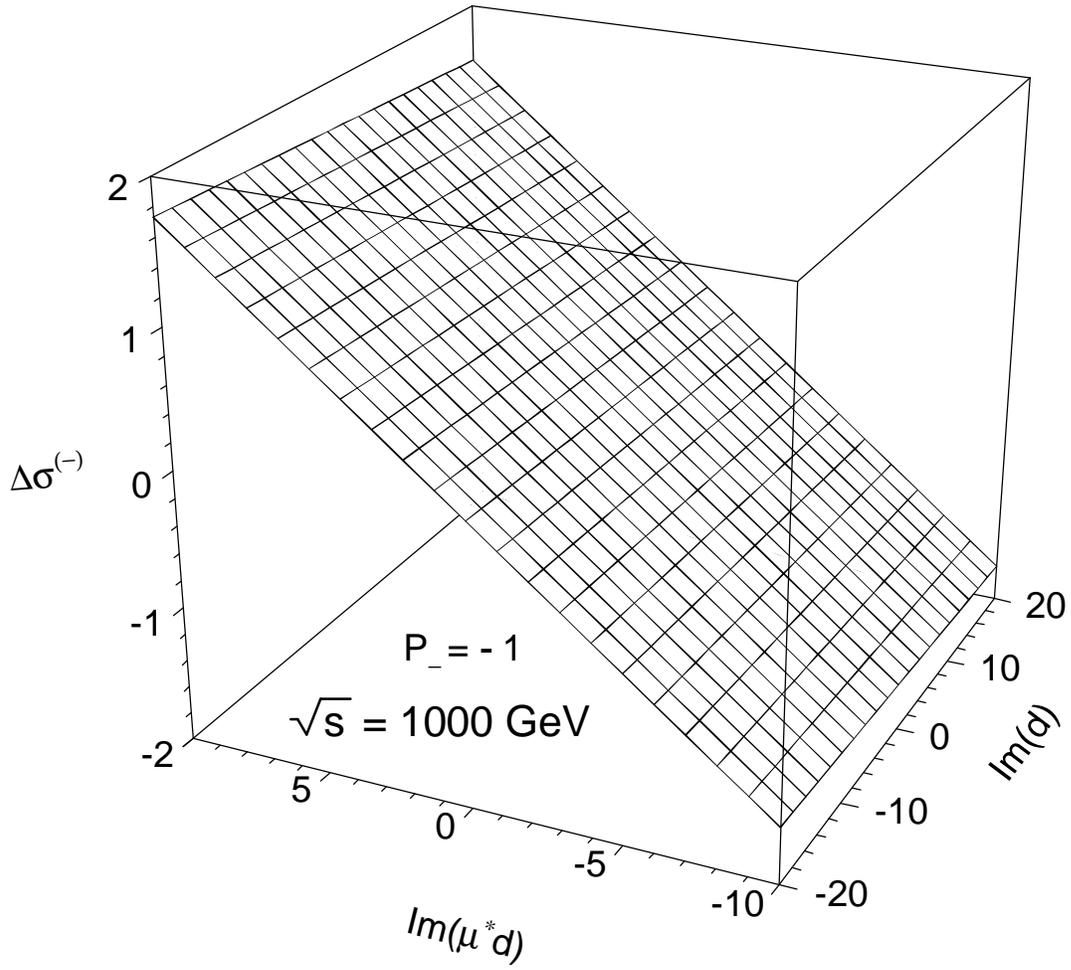
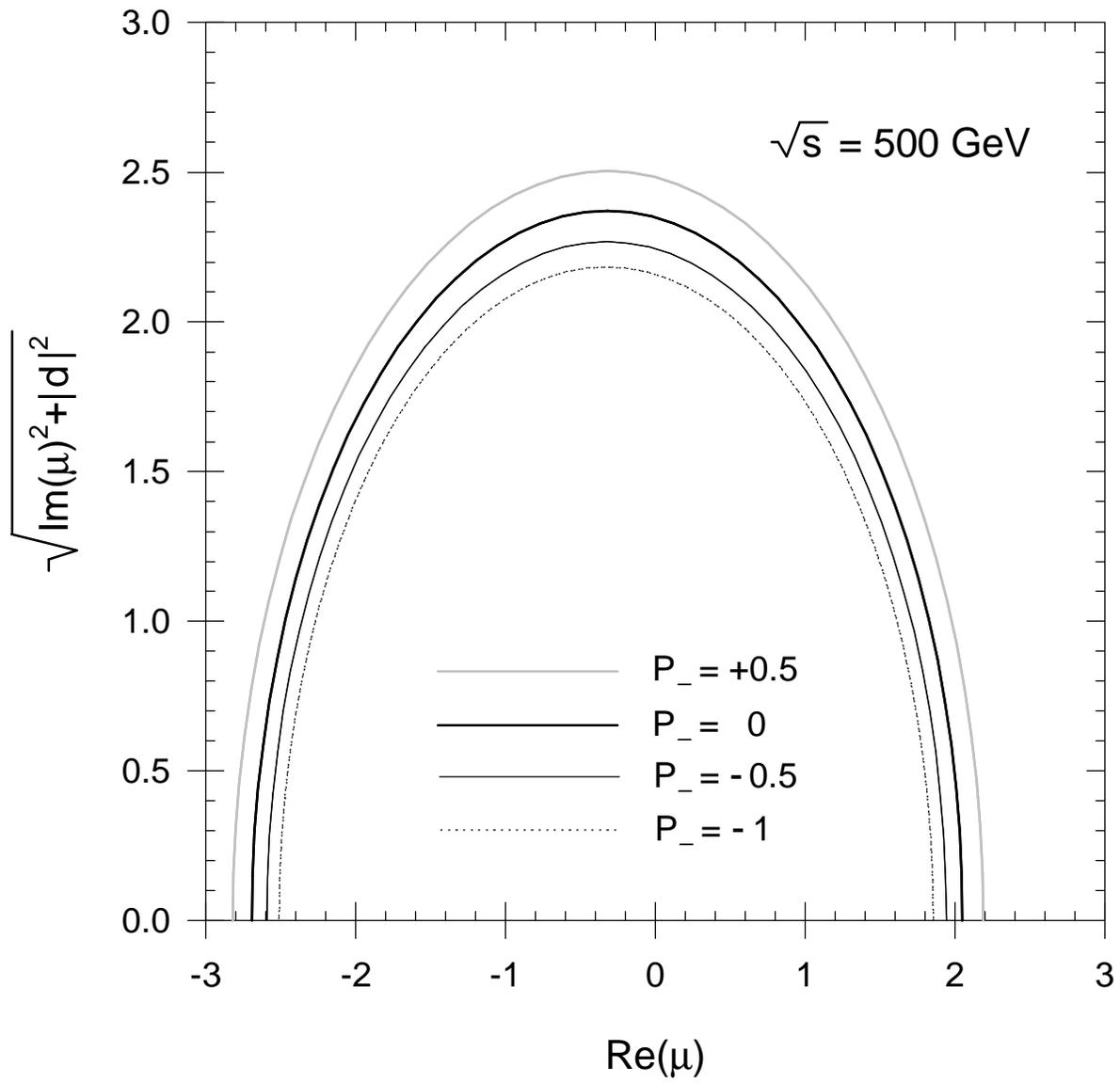
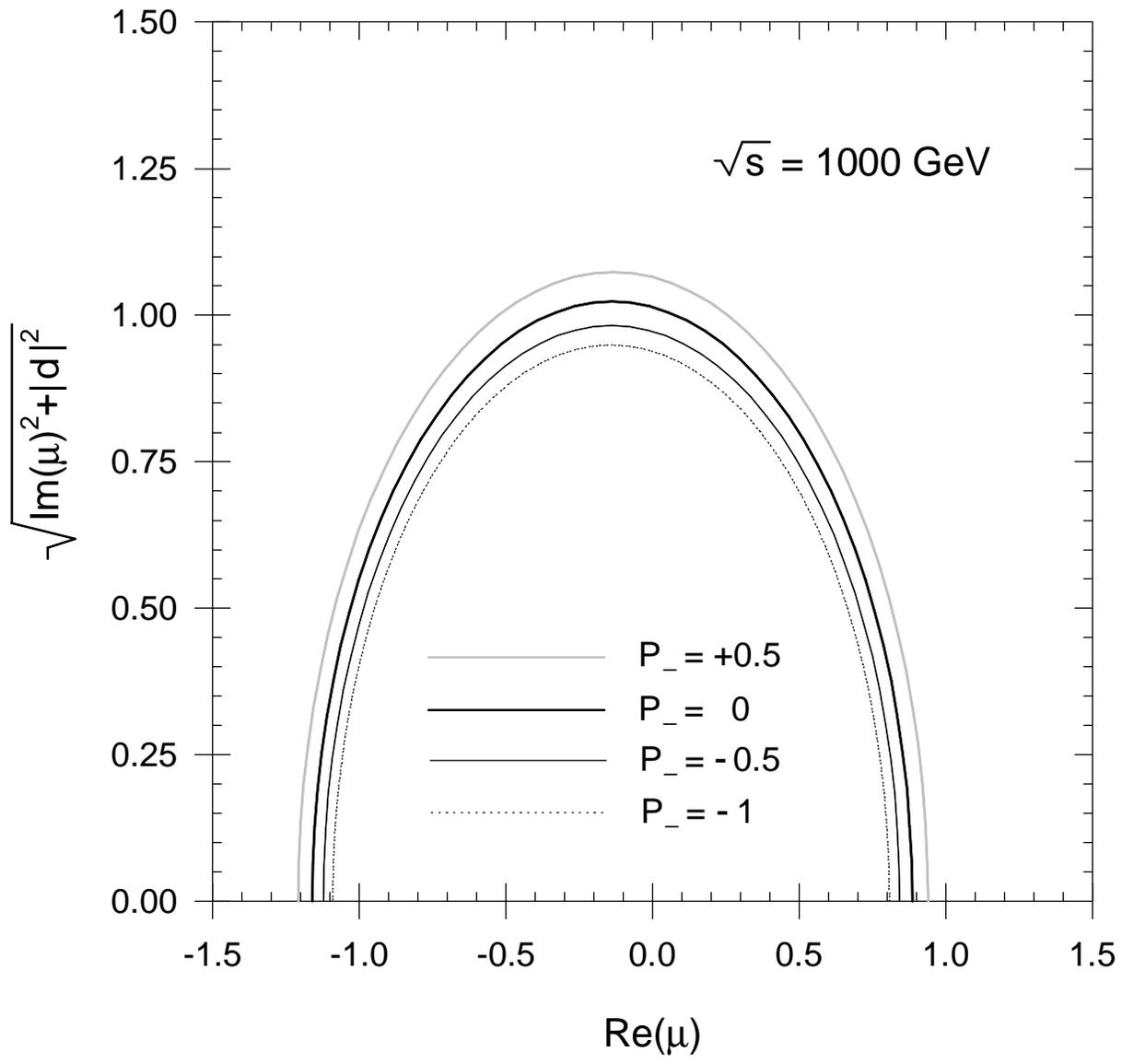
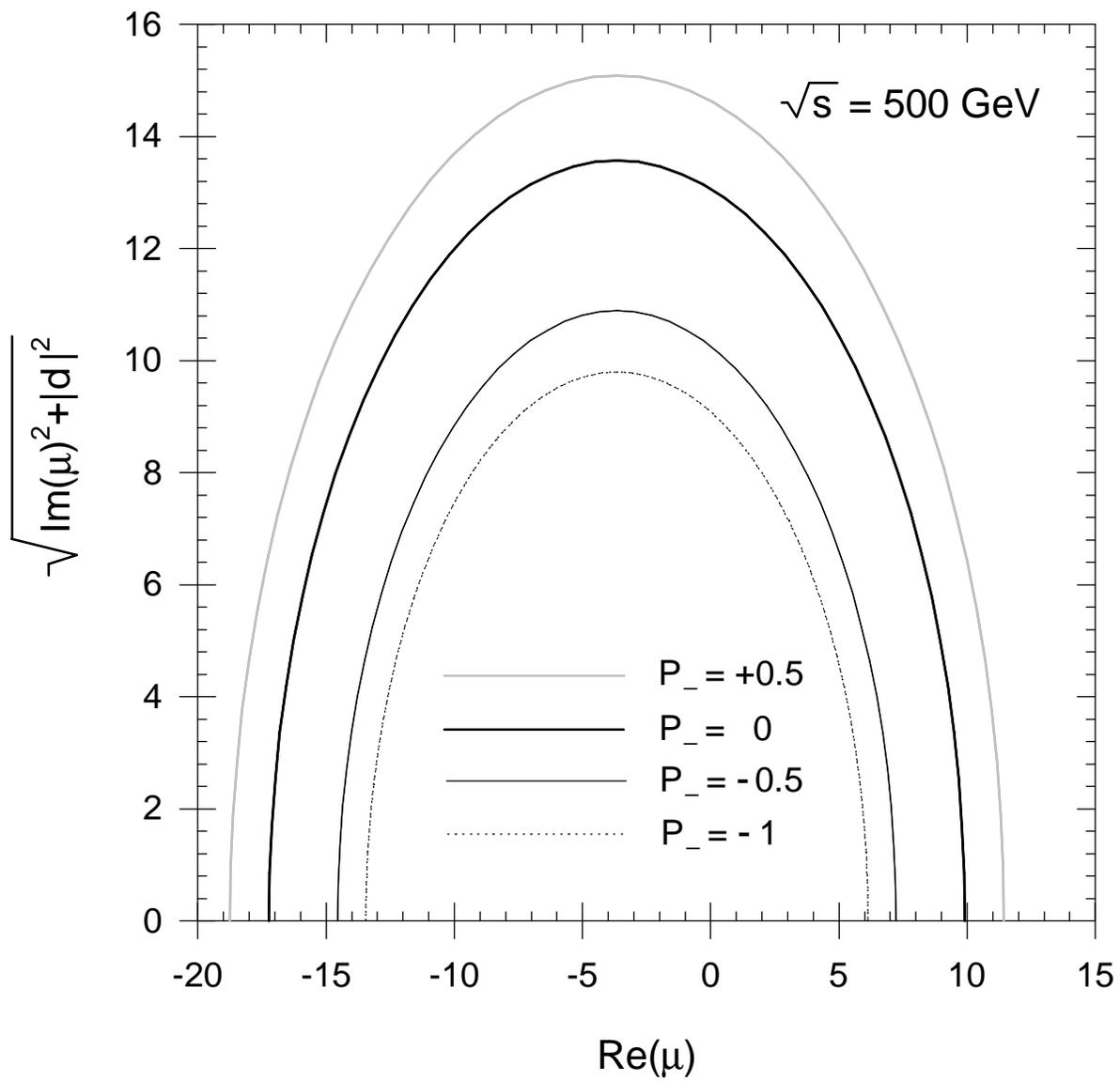
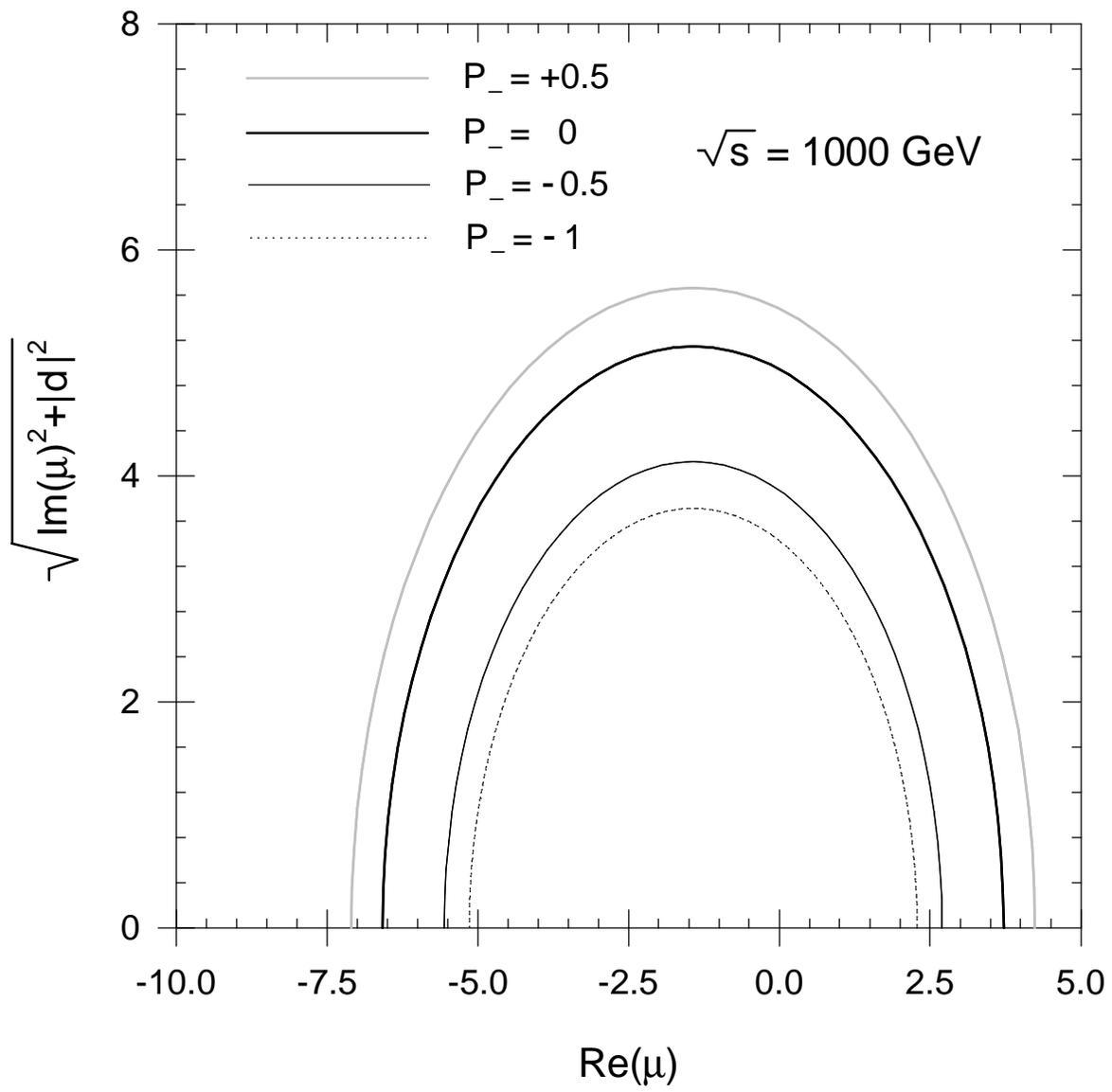
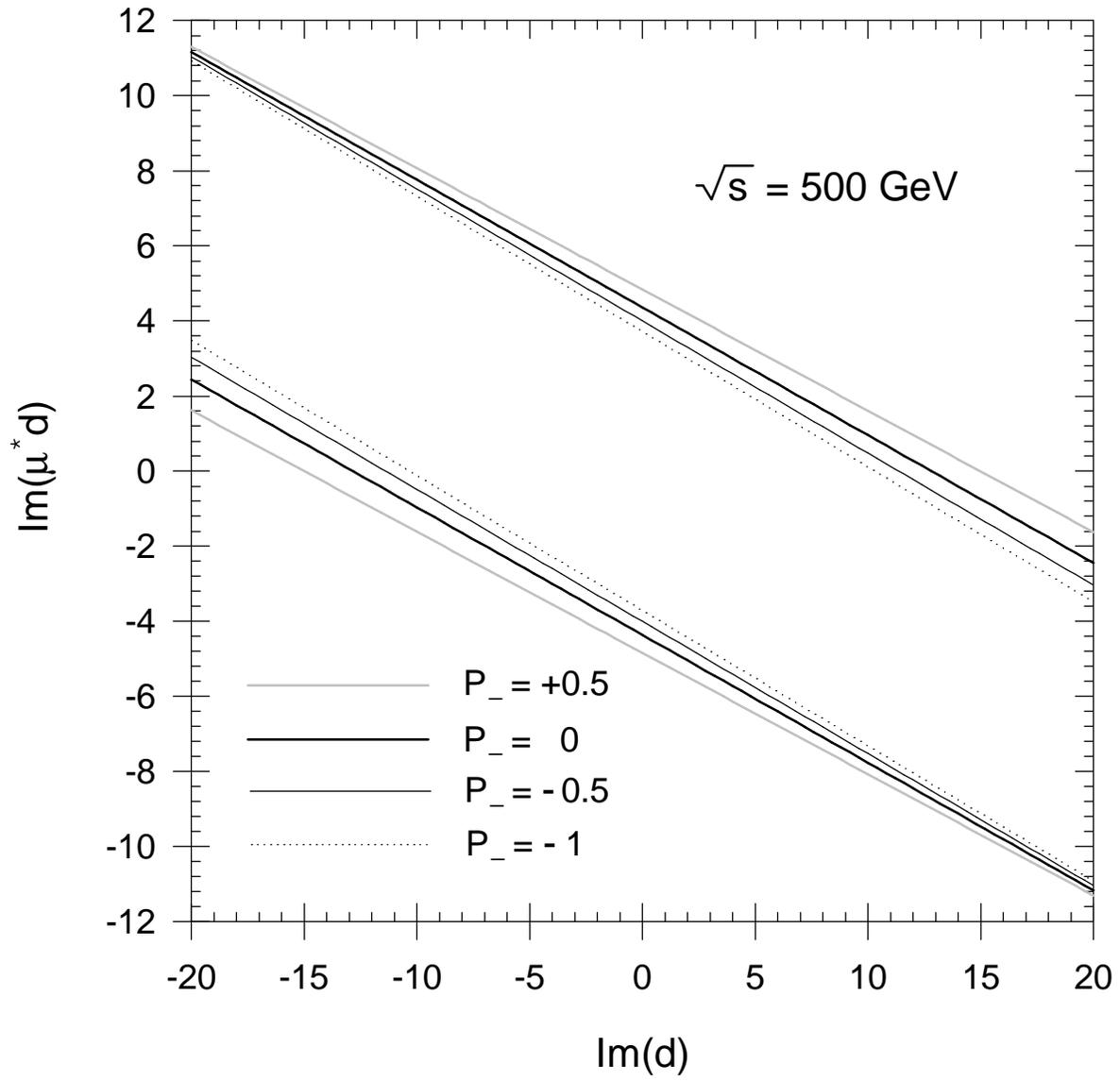
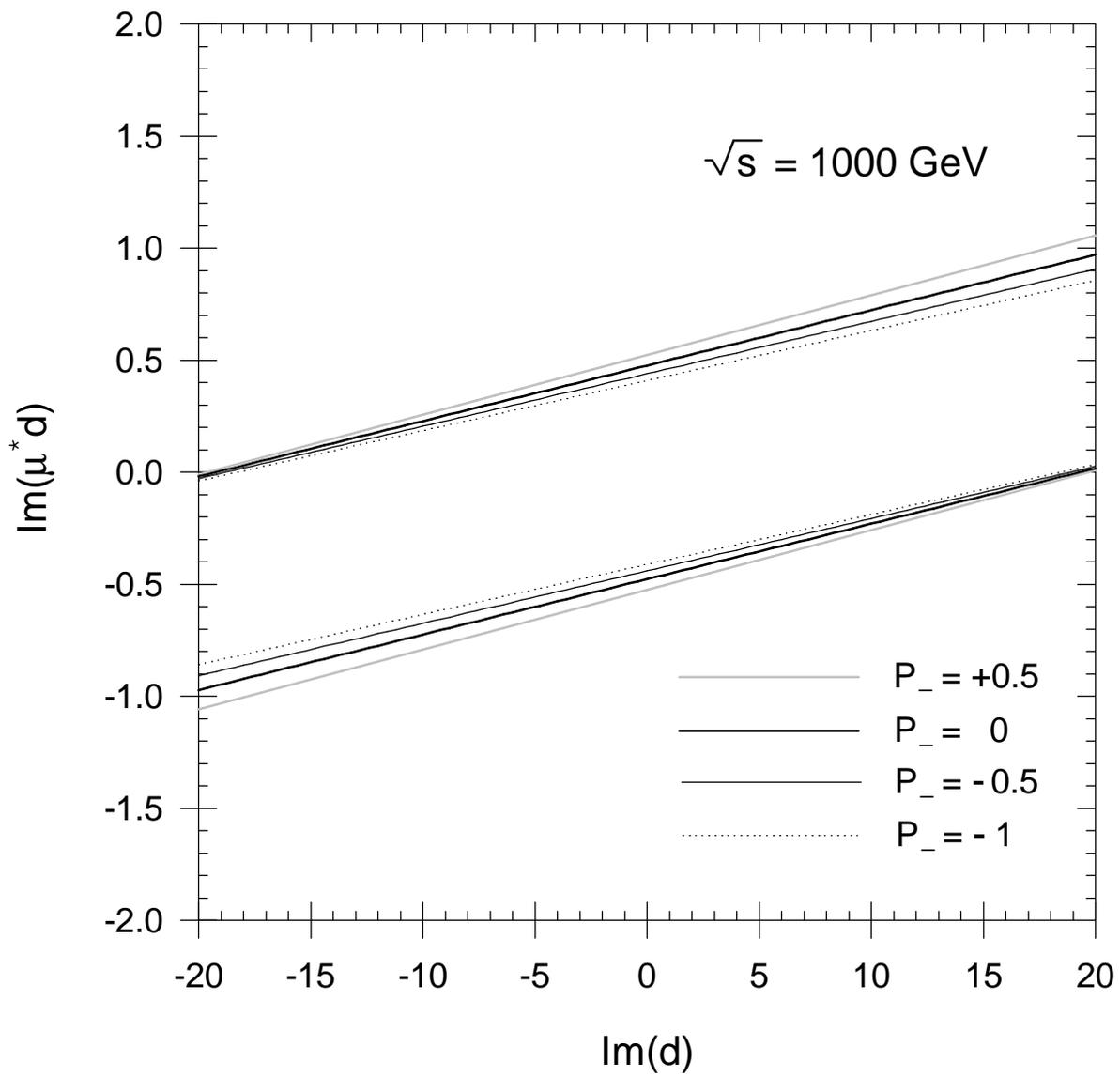
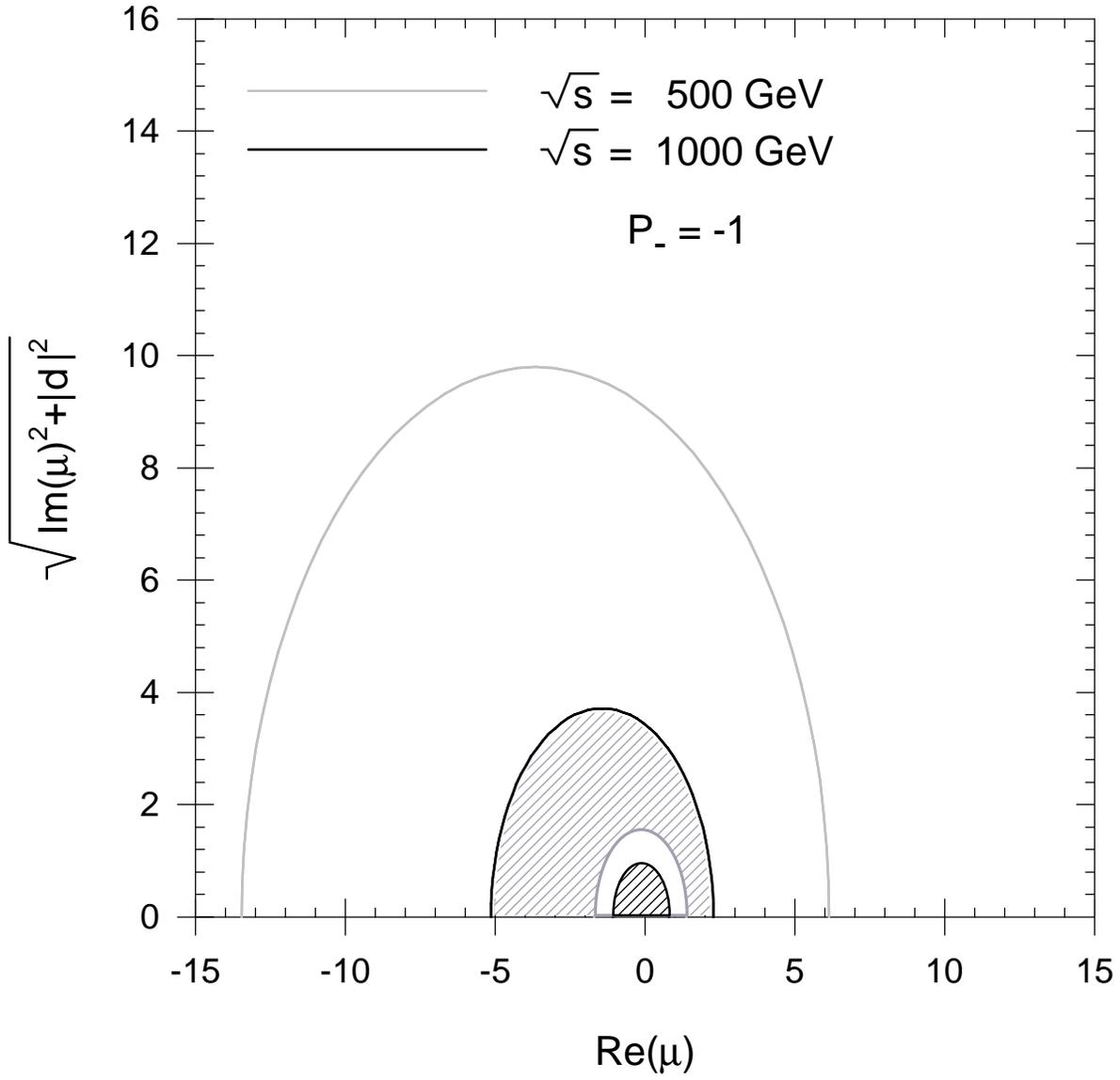
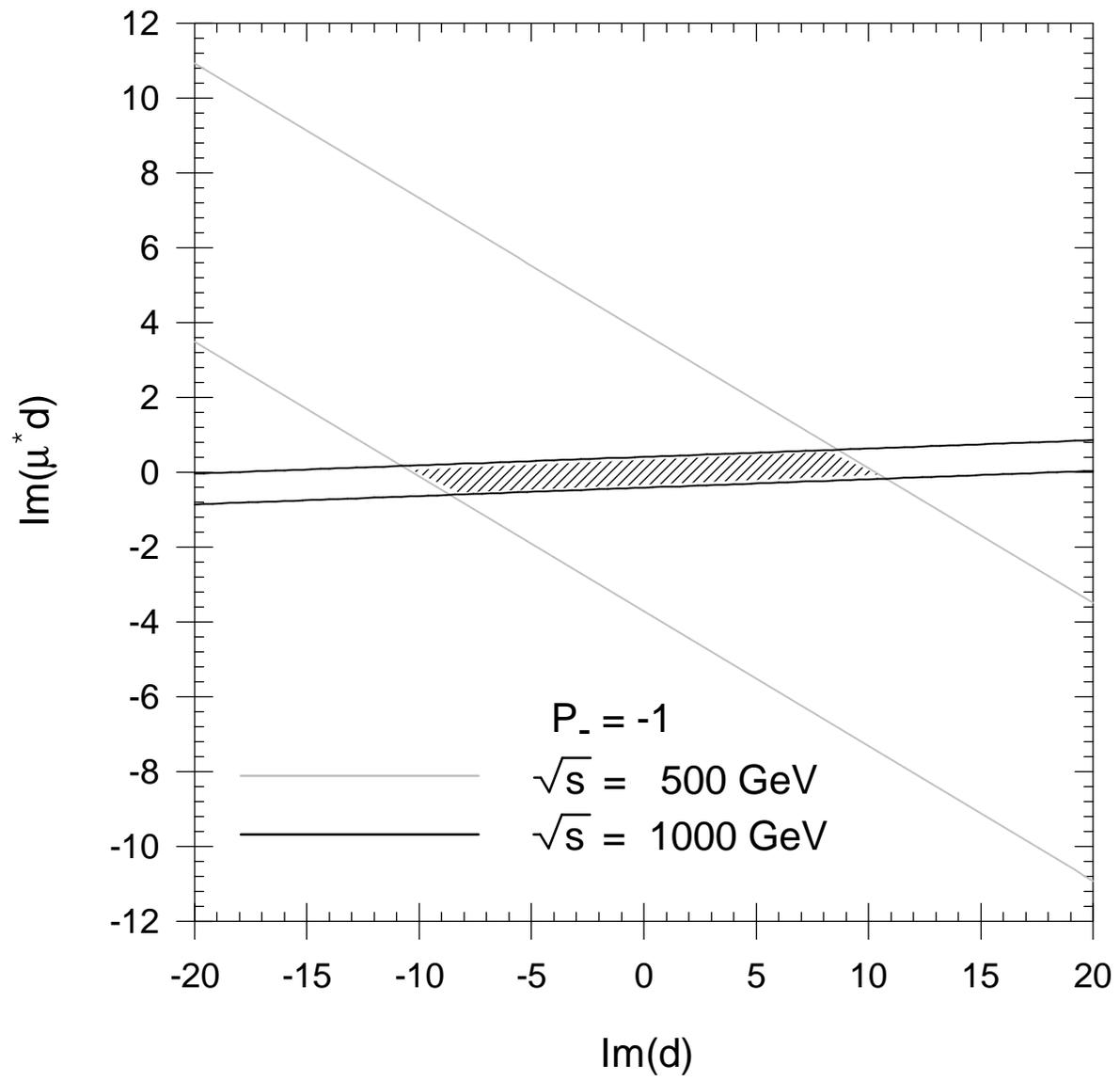
\end{document}